\documentclass[aps,prr,twocolumn,showpacs,superscriptaddress,10p,longbibliography]{revtex4-1}

\usepackage{amsmath}
\usepackage{amssymb}
\usepackage[colorlinks=true,linkcolor=red,citecolor=magenta]{hyperref}
\usepackage[sort&compress]{natbib}
\usepackage{scalerel}
\usepackage[normalem]{ulem}
\usepackage{graphicx}
\usepackage{xcolor}
\usepackage{physics}
\usepackage{dsfont}
\usepackage{mathrsfs}
\usepackage{float}
\usepackage{mathtools}
\usepackage{subfigure}

\newcommand{\pcsadd}{Center for Theoretical Physics of Complex Systems, Institute for Basic Science (IBS), Daejeon 34126, Korea}
\newcommand{\ustadd}{Basic Science Program, Korea University of Science and Technology (UST), Daejeon 34113, Korea}

\newcommand{\mh}{\mathcal{H}}
\newcommand{\mz}{\mathbb{Z}}
\newcommand{\ms}{\mathbb{S}}
\newcommand{\mf}{\mathcal{F}}

\newcommand{\mr}{\mathcal{R}}

\newcommand{\mE}{\mathcal{E}}
\newcommand{\vmE}{\vec{\mathcal{E}}}

\newcommand{\ha}{\hat{a}}
\newcommand{\he}{\hat{e}}
\newcommand{\vlp}{\vec{r}}
\newcommand{\vn}{\vec{n}}
\newcommand{\vnpl}{\vn_\parallel}
\newcommand{\vnpp}{\vn_\perp}

\newcommand{\xa}{x_\alpha}
\newcommand{\ya}{y_\alpha}

\newcommand{\sem}{E}

\begin{document}

\title{Wannier-Stark flatbands in Bravais lattices}

\author{Arindam Mallick}
\email{marindam@ibs.re.kr}
\affiliation{\pcsadd}

\author{Nana Chang}
\affiliation{\pcsadd}
\affiliation{Center for Advanced Quantum Studies, Department of Physics, Beijing Normal University, Beijing 100875, People's Republic of China}

\author{Wulayimu Maimaiti}
\affiliation{\pcsadd}
\affiliation{Department of Physics and Astronomy, Center for Materials Theory, Rutgers University, Piscataway, New Jersey 08854, USA}

\author{Sergej Flach}
\affiliation{\pcsadd}
\affiliation{\ustadd}

\author{Alexei Andreanov}
\email{aalexei@ibs.re.kr}
\affiliation{\pcsadd}
\affiliation{\ustadd}

\date{\today}

\begin{abstract}
    We systematically construct flatbands (FB) for tight-binding models on simple Bravais lattices in space
dimension $d \geq 2$ in the presence of a static uniform DC field. Commensurate DC field directions yield
irreducible Wannier-Stark (WS) bands in perpendicular dimension $d - 1$ with $d$-dimensional eigenfunctions.
The irreducible bands turn into dispersionless flatbands in the absence of nearest neighbor hoppings between
lattice sites in any direction perpendicular to the DC field. The number of commensurate directions which yield
flatbands is of measure one. We arrive at a complete halt of transport, with the DC field prohibiting transport
along the field direction, and the flatbands prohibiting transport in all perpendicular directions as well. The
anisotropic flatband eigenstates are localizing at least factorially (faster than exponential).
\end{abstract}

\maketitle

\section{Introduction}
\label{sec:intro}

Systems with macroscopic degeneracies have been attracting attention due to their high sensitivity to weak perturbations, making them an ideal platform to study effects of perturbations and search for unconventional and exotic phases. One class of systems with macroscopic degeneracies are flatband (FB) systems \cite{lieb1989two,leykam2018artificial,leykam2018perspective,derzhko2015strongly}. FBs are dispersionless energy bands of translational invariant tight-binding networks which occur due to destructive interference. In the past decades flatband systems have been widely studied both theoretically and experimentally with realizations in one-dimensional (1D), two-dimensional (2D) and three-dimensional (3D) setups~\cite{leykam2018artificial,leykam2018perspective,derzhko2015strongly}. Flatbands are by their very definition macroscopically degenerate and highly sensitive to perturbations. Different perturbations lead to different phenomena such as unconventional Anderson localization in presence of disorder~\cite{goda2006inverse,nishino2007flat,chalker2010anderson,leykam2013flat,leykam2017localization,shukla2018disorder}, the appearance of compact breathers in presence of nonlinearity~\cite{maimistov2017on,gligoric2016nonlinear,zegadlo2017single,leykam2013flat,real2018controlled,gligoric2018nonlinear,diliberto2019nonlinear,johansson2015compactification,gligoric2016nonlinear,danieli2018compact}, flatband ferromagnetism in Hubbard model~\cite{mielke1991ferromagnetism,tasaki1994stability,tasaki2008hubbard,derzhko2007low,maksymenko2012flatband}, Landau-Zener-Bloch oscillations in external dc field~\cite{khomeriki2016landau,long2017topological}, enhanced superfluidity in presence of attractive interaction~\cite{peotta2015superfluidity,julku2016geometric,tovmasyan2018preformed}, enhanced superconductivity~\cite{volovik2018graphite}, etc. 

All these phenomena originate from perturbations lifting the macroscopic degeneracy and breaking the destructive interference, which is at the origin of flatbands. Appearance of destructive interference requires either fine-tuning or symmetries, that enforce the interference~\cite{ramachandran2017chiral}. A marked feature of FB models with short-range hopping are strictly compact eigenstates, called \emph{compact localized states} (CLS)~\cite{read2017compactly}. Their presence greatly simplifies the analysis of the FB models~\cite{danieli2020nonlinear,danieli2020quantum}, and can be used as a foundation for their systematic classification~\cite{maimaiti2017compact,maimaiti2019universal,maimaiti2020flatband,maimaiti2020thesis}.

A completely different scenario unfolds when a Bravais lattice is exposed to a dc field, which generates an infinity of $(d-1)$-dimensional bands each supporting eigenstates embedded into $d$-dimensional spaces. In this work we analyze models with applied dc fields that do not need fine-tuning to achieve band flatness, and FB eigenstates that cannot be arranged into CLS. Applying a static field to a 1D tightbinding chain leads to the appearance of a Wannier-Stark (WS) ladder of equidistant eigenenergies for the spectrum of the chain, with all eigenstates being localized and the dynamics of observables in general displaying time-periodic Bloch oscillations~\cite{maksimov2015wannierI}. The quest for nontrivial states in higher dimensional lattices with magnetic and electric fields resulted in the observation of flatbands for a square lattice for certain electric field directions~\cite{nakanishi1995two}. Similar dispersionless features were later identified for a rectangular lattice~\cite{keck2002infinite,bulgakov2014induced}. Here we present a systematic construction of WS flatbands for the five two-dimensional Bravais lattices: Oblique, rectangular, centered rectangular, triangular, and square lattices. We obtain the dependence of the band structure of the static field direction, analyze the localization properties of the flatband eigenstates, and demonstrate that, unlike conventional FB systems which host CLS, WS flatband states cannot host CLS and are at least factorially localized instead. We then generalize these results to higher lattice dimensions and discuss the impact of longer-range hoppings.

The paper is organized as follows: Sec.~\ref{sec:model} introduces the model for 2D lattices and the definitions that are used later in Sec.~\ref{sec:spectrum} to derive the band structure of the models for different directions of the field. Conditions for the bands flatness are also discussed in the same Sec.~\ref{sec:spectrum}, while the eigenstates are analyzed in Sec.~\ref{sec:eigs}. We extend our analysis to higher lattice dimensions in 
Sec.~\ref{higher_dim}. We conclude with a summary and open issues.

\section{Setting the stage}
\label{sec:model}

\begin{figure}
    \includegraphics[width=0.99\columnwidth]{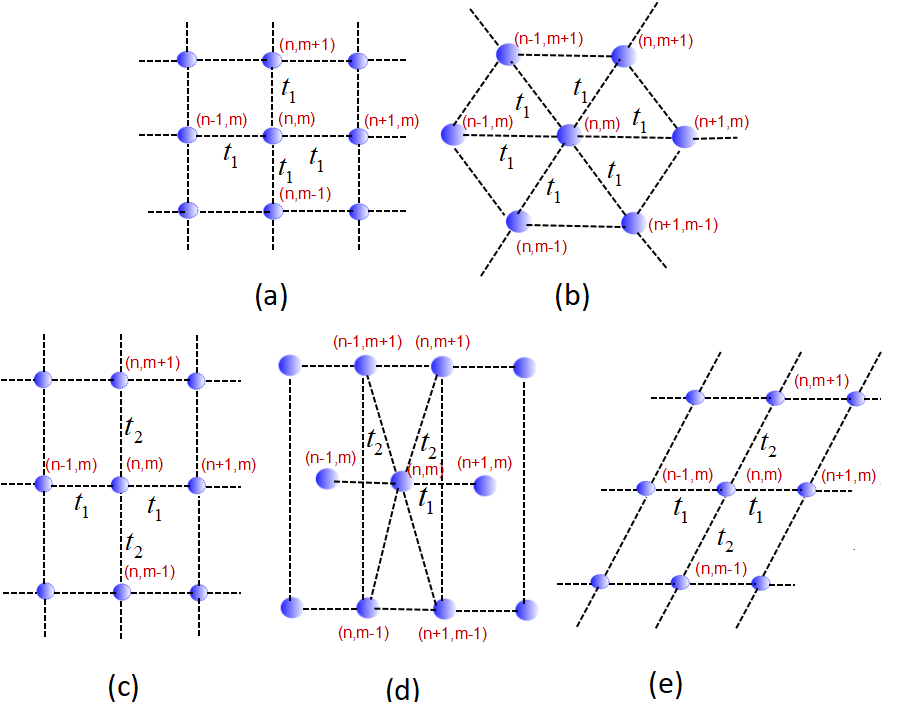}
    \caption{Schematics of the five two-dimensional Bravais lattices with nearest neighbor hoppings: (a) square; (b) triangular; (c) rectangular; (d) centered rectangular; (e) oblique. $t_1$, $t_2$ are the hopping strengths.}
    \label{fig:bravais}
\end{figure}

We consider a 2D Bravais lattice (see Fig.~\ref{fig:bravais}):
\begin{align}
    \vlp_{nm} = n \ha_0 + m \ha_1,
\end{align}
where $\ha_{0,1}$ are the lattice basis vectors and $(n,m)$ are the translation indices of the lattice point. The vectors need not be orthogonal, for example in the case of the triangular lattice, and can be expressed in terms of Euclidean basis vectors ($\he_0$, $\he_1$) as
\begin{align}
    \ha_0 = \gamma_0 \he_0,~ \ha_1 = \gamma_1(\cos\theta\, \he_0 + \sin\theta\, \he_1).
\end{align} 
Here $0 < \theta \leq \pi/2$ is a tilting angle of the axis, and $\gamma_{0,1}$ are the lengths of the basis vectors of the lattice unit cell. For square and triangular lattice $\gamma_0 = \gamma_1$. For all other cases (oblique, rectangular and centered rectangular) they are not equal. For rectangular and square lattices $\theta = \frac{\pi}{2}$, for triangular lattice $\theta = \frac{\pi}{3}$, and for oblique and centered rectangular lattices $\theta$ can take any value except $\frac{\pi}{2}$ and $\frac{\pi}{3}$. 

Next we define a tight-binding Hamiltonian on the lattice in the presence of a static dc field $\vmE$:
\begin{align}
    \mh =  \sum_{n,m} \Big[ - \sum_{l,j} t_{lj} \ket{n-l, m-j}\bra{n, m} \notag\\
    \label{eq:tb-ham}
    +~\vmE \cdot \vlp_{nm} \ket{n,m}\bra{n,m} \Big],
\end{align}
which acts on the Hilbert space spanned by the basis vectors $\{\ket{n,m}$: $(n,m)$ $\in$ $\mz \times \mz\}$. The indices $l, j$ denote the hopping range.

Our main goal is to compute and analyze the spectrum of this Hamiltonian. Because of the presence of the dc field, discrete translation invariance is in general broken and we do not expect any eigenenergy band structure. However the character of the spectrum depends on the direction of the dc field, which we set to 
\begin{align}
    \vmE = F \vnpl = F \frac{\left(x \ha_0 + y \ha_1\right)}{|x \ha_0 + y \ha_1|}.
\end{align}
Here $F$ is the strength of the field and $x,y\in\mathbb{R}$. Then, parallel and perpendicular directions to the field are encoded by the respective unit vectors
\begin{align}
    & \vnpl = \frac{(x \gamma_0 + y \gamma_1 \cos\theta)\he_0 + (y \gamma_1 \sin\theta) \he_1}{\sqrt{\gamma^2_0 x^2 +   \gamma^2_1 y^2 + 2 \gamma_0 \gamma_1 x y \cos\theta}},\notag\\
    & \vnpp = \frac{(y \gamma_1 \sin\theta) \he_0 - (x \gamma_0 + y \gamma_1 \cos\theta) \he_1}{\sqrt{\gamma^2_0 x^2 +   \gamma^2_1 y^2 + 2 \gamma_0 \gamma_1 x y \cos\theta}}~.
\end{align} 
We choose the direction of the field such that $\vnpp$ is parallel to one of the lattice vectors $p_1\ha_0 + p_2\ha_1$ with $p_1, p_2 \in \mz$. This ensures that the translational invariance of the lattice persists in the perpendicular direction $\vnpp$, albeit with a different lattice spacing than the original lattice. In what follows we coin such field directions as \emph{commensurate directions}. They ensure the existence of a 1D band structure in the spectrum of the model. Commensurate field directions constrain the possible values of $x,y$ and $\theta$ to either of the two possibilities
\begin{gather}
    \begin{cases}
        |x\gamma_0| = |y\gamma_1| ~\text{and}~\frac{x}{y}, \frac{\gamma_0}{\gamma_1} ~\text{are rational}\\
        \cos\theta = -\frac{p_1 x \gamma^2_0 + p_2 y \gamma_1^2} {(p_2  x + p_1 y)\gamma_0 \gamma_1}
    \end{cases},
    \label{eq:comm_condition}
\end{gather}
as discussed in Appendix~\ref{app:derivation_comm}.
\begin{figure}
    \includegraphics[width=1.05\columnwidth]{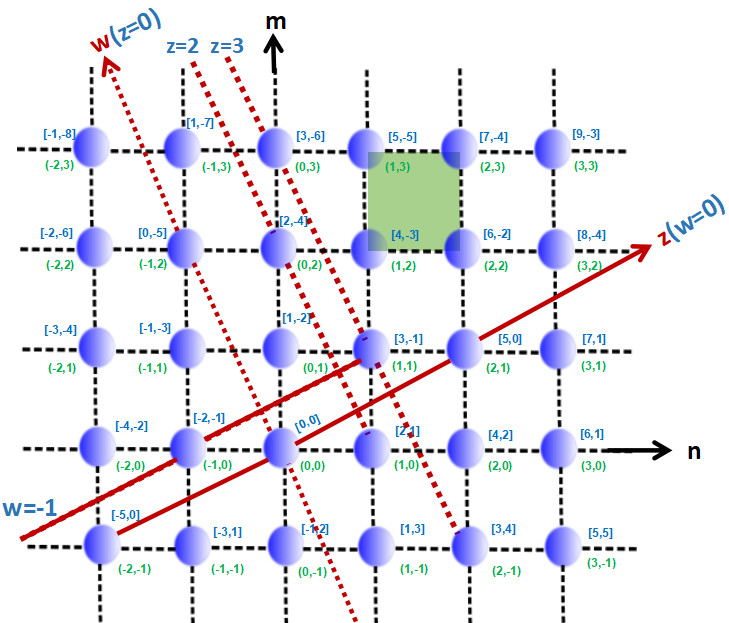}
    \caption{Pictorial representation of the two coordinate systems for the square lattice: $(n,m)$ shown as numbers within parenthesis (green color), and $(z,w)$ shown as numbers within square brackets (blue color). The unit cell is the green shaded area formed by the vertices: $\{n = 1,2; m = 2,3\}$. The direction of the dc field is $(x,y) = (2,1)$ along the $z$ axis. The red solid and red dashed lines represent, respectively, the lines of constant $w$ and $z$.}
    \label{fig:unit-cell}
\end{figure}

In order to diagonalize the Hamiltonian [Eq.~\eqref{eq:tb-ham}], we exploit the partial translation invariance and introduce a new rotated coordinate system (see Fig.~\ref{fig:unit-cell}): the $z$ coordinate along the dc field, and the $w$ coordinate perpendicular to the dc field that we define as follows
\begin{subequations}
    \begin{align}
        z &= \alpha\sqrt{\gamma^2_0 x^2 + \gamma^2_1 y^2 + 2 \gamma_0 \gamma_1 x y \cos\theta}~\vlp_{nm} \cdot \vnpl \notag\\
        &= n x_\alpha + m y_\alpha, \\
        w &= \beta\sqrt{\gamma^2_0 x^2 + \gamma^2_1 y^2 + 2 \gamma_0 \gamma_1 x y \cos\theta}~\vlp_{nm} \cdot \vnpp \notag\\
        &= n y - m x,
    \end{align}
    \label{eq:zw-def}
\end{subequations}
with
\begin{subequations}
    \begin{align}
        \xa = \alpha (x \gamma_0^2 + y \gamma_0 \gamma_1 \cos\theta), \\
        \ya = \alpha (x \gamma_0 \gamma_1 \cos\theta + \gamma_1^2 y),
    \end{align}
\end{subequations}
and $\beta = (\gamma_0 \gamma_1 \sin\theta)^{-1}$. For any commensurate field direction there exists a value of the coefficient $\alpha$ such that both $\xa, \ya$ become mutually prime integers [as can be directly verified by computing the ratio $\ya/\xa$ using the conditions~\eqref{eq:comm_condition}, see Appendix~\ref{app:para}]. Then we can parametrize the new coordinates as
\begin{align}
    (z, w) = (z, w_0(z) + \eta(x \xa + y \ya)),~\quad z,\eta \in \mz.
    \label{eq:z_w_val}
\end{align}
with a function $w_0(z)$, see Appendix~\ref{app:para} for details. Finally we note that for any coprime $(x,y)$ (as happens for any dc field commensurate direction on a square lattice) the distances $\Delta_z$ and $\Delta_w$ in real space covered by changing respectively, $z$ to $z+1$ and $\eta$ to $\eta+1$, amount to
\begin{align}
    & \Delta_z = \sqrt{\gamma^2_0 x^2 + \gamma^2_1 y^2 + 2 \gamma_0 \gamma_1 x y \cos\theta}, \notag \\
    & \Delta_w = \frac{\alpha}{\beta}\sqrt{\gamma^2_0 x^2 + \gamma^2_1 y^2 + 2 \gamma_0 \gamma_1 x y \cos\theta}. 
    \label{eq:DZDW}
\end{align}

\section{The 2D spectrum}
\label{sec:spectrum}

The convenience of the $z,w$ basis is that now the partial translation invariance of the Hamiltonian $\mh$ [Eq.~\eqref{eq:tb-ham}] in the direction perpendicular to the field is made explicit:
\begin{align}
    \mh = &\sum_{(z,w)} \Big(\mf z \ket{z,w}\bra{z,w} \notag \\
    & -\sum_{l,j} t_{lj} \ket{z - l\xa - j\ya, w - ly + jx}\bra{z, w}\Big).
    \label{eq:tight_bnew_ham}
\end{align} 
In the above we have also defined
\begin{align}
    \mf = \frac{F}{\alpha \sqrt{\gamma^2_0 x^2 + \gamma^2_1 y^2 + 2 \gamma_0 \gamma_1 x y \cos\theta}}.
\end{align}
The Hamiltonian is invariant under the shifts $w \to w'$, but not under the shifts $z \to z'$. Therefore we can apply the Bloch's theorem in the direction perpendicular to the field. For that we define a complete orthonormal set of basis vectors
\begin{align}
    \label{eq:basis}
    \ket{\phi(z, k)} = \ket{z} \otimes \sum_{\eta \in \mz} e^{i k \eta}\ket{w_0(z) + \eta(x x_\alpha + y y_\alpha)}, \\
    \bra{\phi(z',k')}\ket{\phi(z,k)} = 2 \pi \delta_{z,z'}\delta(k-k'), \notag
\end{align}
where $z \in \mz$ and $k \in [0, 2\pi)$, the quasi-momentum in the direction perpendicular to the field. Then the action of the Hamiltonian $\mh$ in Eq.~\eqref{eq:tb-ham} on a basis vector is given by
\begin{align}
    & \mh\ket{\phi(z, k)} = \mf z \ket{\phi(z, k)}  \notag \\
    \label{eq:basis_oper}
    & -\sum_{l,j} t_{lj}~e^{ik \epsilon_{lj}} \ket{\phi(z - l \xa - j \ya, k)}, \\
    & \epsilon_{lj} = -(j\tau_2 - l \tau_1), \notag
\end{align}
where we have used Eq.~\eqref{eq:ident_rel} to simplify the expression. The Hamiltonian does not couple different values of $k$ and the basis states [Eq.~\eqref{eq:basis}] for fixed $k$ form invariant subspaces of the Hamiltonian. Therefore we can use the following ansatz for the eigenstates of $\mh$:
\begin{align}
    \ket{\psi_\sem(k)} = \sum_z \psi_\sem(z,k) \ket{\phi(z, k)}
\end{align}
which correspond to an eigenvalue $E$. The eigenvalue is a function of the quasimomentum $k$:
\begin{align}
    H\ket{\psi_\sem(k)} = E\ket{\psi_\sem(k)}.
    \label{eq:eigen_eq}
\end{align}

\subsection{Generating function method}

Using Eq.~\eqref{eq:basis_oper} we write the eigenproblem
\begin{align}
    (\mf z - E) \psi_\sem(z, k) = \sum_{l,j} t_{lj}~e^{ik\epsilon_{lj}}
    \psi_\sem(z + l \xa + j \ya, k),
    \label{eq:reduce_equ_motion}
\end{align}
and solve it by introducing a generating function 
\begin{align}
    \label{eq:GF}
    g_\sem(q, k) = \sum_{z \in \mz} e^{-iqz} \psi_\sem(z, k), \\ 
    \label{eq:PC}    
    g_\sem(q+2\pi, k) = g_\sem(q, k+2\pi) = g_\sem(q, k).
\end{align}
We note that the generating function $g_\sem(q,k)$ is nothing but the Fourier transformed $ \psi_\sem(z, k)$ from $z$-space to $q$-space. The eigenproblem, Eq.~\eqref{eq:reduce_equ_motion}, transforms into an ordinary differential equation
\begin{align}
    \label{eq:eig_g_equation}
    i \mf \frac{\partial}{\partial q} g_\sem(q, k) - E g_\sem(q, k) = \notag\\
    \sum_{l,j} t_{lj} e^{ik\epsilon_{lj}} e^{i q(l \xa + j \ya)} g_\sem(q, k). 
\end{align} 
This equation can be solved for any pair of $(E,k)$. The solutions are \emph{not} periodic functions in $q$ for generic $E, k$. The periodicity in the $q$ requirement from Eq.~\eqref{eq:PC} imposes a dispersion relation of WS bands between $E$ and $k$.

We assume that the hopping networks on each of the five Bravais lattices respect \emph{inversion symmetry}: 
\begin{align}
	\label{eq:inv_sym}
    t_{l,j}  = t_{-l,-j}\qquad\forall~j, l. 
\end{align}
This assumption is made for convenience only, the conclusions presented below hold as well if the symmetry is broken. Then,  defining $\lambda_\sem = E/\mf,~ s_{lj} = t_{lj}/\mf$ we arrive at
\begin{align}
    \frac{\partial \ln[g_\sem(q, k)]}{\partial q} = -i\lambda_\sem 
    - i\sum_{l, j} s_{lj}\cos[k\epsilon_{lj} + q(l \xa + j \ya)].   
    \label{eq:diff_g}
\end{align}
This differential equation can be easily integrated, and we discuss in the next section the possible solutions.

\begin{widetext}

\subsection{Flatbands and Dispersive Bands}
The set of commensurate field directions splits into a subset of flatband directions and the complementary subset of dispersive directions. All flatband directions satisfy the condition
\begin{equation}
    l \xa + j \ya \neq 0 \;,\; {\rm for \; all} \; l, j \;.
    \label{eq:fbdirection}
\end{equation}
This condition is equivalent to requesting the absence of any direct hopping connection between two lattice sites perpendicular to the chosen commensurate field direction in Eq.~\eqref{eq:tight_bnew_ham}. Indeed, for this case the solution of Eq.~\eqref{eq:diff_g} reads as 
\begin{align}
g_\sem(q, k) = A(k) e^{-i\lambda_\sem q}
\exp\left(-i\sum\limits_{l, j} \frac{s_{lj}}{l \xa + j \ya} \sin[k\epsilon_{lj} 
+  q(l \xa + j \ya)]\right). \label{eq:flat_g} 
\end{align}
Enforcing $2\pi$ periodicity of $g_\sem$ in $q$ results in
\begin{align}
    \lambda_\sem = a\in\mz \Rightarrow E = \mf a,
    \label{eq:flat_ge}
\end{align}
i.e., all bands are flat, equidistant, and labeled by an integer index $a$.
We note that the generating function remains periodic in $q$ and the bands remain flat even in the absence of the inversion symmetry~\eqref{eq:inv_sym}.

For a dispersive field direction $l \xa + j \ya = 0$ for at least one pair $(l=l',j=j')$, it follows
\begin{align}
       g_\sem(q, k) =  A(k)
    \exp\left(- i\{\lambda_\sem + \sum_{l', j'} s_{l'j'} \cos[k\epsilon_{l'j'}]\} q\right) \exp\left(-i\sum_{(l, j) \neq (l', j') } \frac{s_{lj}}{l \xa + j \ya} \sin[k\epsilon_{lj} + q(l \xa + j \ya)]\right).  \label{eq:disperse_g}
\end{align}
Again enforcing the $2\pi$ periodicity of $g_\sem$ in $q$ we get:
\begin{align}
     \lambda_\sem + \sum_{l', j'} s_{l'j'} \cos[k\epsilon_{l'j'}] = a\in\mz ~~ \Rightarrow E = \mf a - \sum_{l', j'}  t_{l'j'} \cos[k\epsilon_{l'j'}]. 
    \label{eq:disperse_ge}
\end{align}
All bands turn dispersive, equidistant and labeled by an integer index $a$. For both cases the band gap between two consecutive energy bands is equal to $\mf$ for any $k$. The prefactor $A(k)$ is periodic in $k$: $A(k) = A(k + 2\pi)$.

It is instructive to observe that the dc field strength enters completely additive in the dispersion relation~\eqref{eq:disperse_ge}, leaving the irreducible $E(k)$ dependence invariant. Also, for any short-range network the number of flatband directions is infinite, while the number of dispersive ones is always finite and equal to the number of hopping connections on the network. Adding more hoppings to a given Bravais lattice network will add more dispersive field directions on the expense of the flatband directions. The limiting case of connecting all sites with all (even though the hopping strength may decrease in a suitable way with increasing distance between sites) will eliminate all flatband directions and leave us with dispersive field directions only.
\end{widetext}
\section{Localization of 2D flatband eigenstates}
\label{sec:eigs}
We next turn to the analysis of the eigenstates of the WS flatbands. Flatbands enjoy macroscopic degeneracy as there is no unique choice of the eigenstate basis. Eigenstates of flatbands in short-range translationally invariant Hamiltonians can be typically arranged into compact localized states \cite{read2017compactly,leykam2018artificial}. We note that in such translationally invariant cases the number of eigenstates of one flatband equals the embedding space dimension of its eigenvectors. For WS flatbands the situation differs, as the embedding space dimension for eigenvectors is infinitely larger than the number of eigenstates of one flatband. It appears impossible to assemble a linear combination of WS flatband eigenstates, which turns compact in real space. Indeed, let us assume that a compact localized state does exist. Then, the generating function $g(q,k)$ can be expanded into a double Fourier series with a finite number of components in both $q$ and $k$. This contradicts the general solution obtained in Eq.~\eqref{eq:flat_g}. Therefore, compact localized states are ruled out---see Appendix~\ref{app:cls_nonproof} for details. What is then the best localization which can be achieved with WS flatbands?

Let us attempt to identify the most localized eigenstates. The eigenstates are extracted from the generating function $g_\sem(q, k)$ for a fixed band $a$. We set $a=0$ and $E=0$ without loss of generality. Using the property of the Bessel function of first kind 
\begin{align}
    e^{i \mu \sin \xi}  = \sum_{\nu \in \mz} J_\nu(\mu) e^{i \nu \xi},
    \label{eq:bessel_pro}
\end{align}
we can express the generating function as
\begin{align}
    g(q, k) =  A(k) \prod_{(l, j) \in \mr} \sum_{\nu_{(l,j)}\in \mz} J_{\nu_{(l,j)}}\left(- \frac{2 s_{lj}}{l \xa + j \ya} \right) \notag\\
    \times e^{i \nu_{(l,j)}[k \epsilon_{lj} + q l \xa + q j \ya]} \notag\\
    = \sum_{\{\nu \in \mz\}}  \left[\prod_{(l, j) \in \mr} J_{\nu_{(l,j)}}\left(- \frac{2 s_{lj}}{l \xa + j \ya} \right) \right] \notag\\
    \times e^{i\sum_{(l, j) \in \mr} \nu_{(l,j)}[k \epsilon_{lj} + q l \xa + q j \ya]},
\end{align}
where $\mr$ denotes the set of the hoppings $l,j$ up to the inversion/reflection symmetry~\eqref{eq:inv_sym}, e.g., for nearest neighbor (n.n.)~hopping these sets are $\{(1,0), (0, 1), (1, -1)\}$ and $\{(1,0), (0, 1)\}$, respectively, in the case of triangular or centered rectangular lattices and all the other lattices. In this case $\nu$ denotes the set of all possible integers $\nu_{(l,j)}$ for all the hoppings $(l,j) \in \mr$. Next we write the flatband basis states in the $z,k$ representation  
\begin{align}
    \ket{\psi(k)} = A(k) \sum_{z \in \mz} \frac{1}{2\pi} \int_{q = 0}^{2\pi} g_\sem(q,k) e^{iqz} dq \ket{\phi(z, k)}.
\end{align}
It follows 
\begin{align}
    \ket{\psi(k)} =  & A(k) \sum_{z \in \mz} \sum_{\{\nu \in \mz\}} 
    \left[\prod_{(l, j) \in \mr} J_{\nu_{(l,j)}}\left(- \frac{2 s_{lj}}{l \xa + j \ya} \right) \right] \notag\\
    & \times\exp\left(i k \sum_{(l, j) \in \mr} \nu_{(l,j)} \epsilon_{lj}\right) \ket{\phi(z, k)},
    \label{eq:z_decay}
\end{align}
subject to the constraint
\begin{align}
    \sum_{(l, j) \in \mr} \nu_{_{(l,j)}}(l \xa + j \ya) = -z~.
    \label{eq:z_decay_cons}
\end{align} 
Because the WS band is flat any linear combination of the basis vectors~\eqref{eq:z_decay}
\begin{align}
    \ket{\Phi} = \int_{k = 0}^{2\pi} c_k \ket{\psi(k)} dk
\end{align}
is an eigenstate of the Hamiltonian $\mh$~\eqref{eq:tb-ham}. Let us choose $c_k=[2 \pi A(k)]^{-1}$ to remove the normalization factor. It follows
\begin{align}
    & \ket{\Phi} =  \sum_{z \in \mz} \sum_{\eta \in \mz} \prod_{(l, j) \in \mr} \left[\sum_{\nu_{(l,j)} \in \mz} J_{\nu_{(l,j)}}\left(- \frac{2 s_{lj}}{l \xa + j \ya} \right) \right]\notag\\
    &~~~~~~~~~~~~~~~~~~~~~~~\ket{z} \otimes \ket{w_0(z) + \eta(x \xa + y \ya)},
    \label{eq:phi_eigv}
\end{align}
subject to the constraints
\begin{align}
    &~\sum_{(l, j) \in \mr} \nu_{(l,j)}(l \xa + j \ya) = -z,\sum_{(l, j) \in \mr} \nu_{(l,j)} \epsilon_{lj} = - \eta.
    \label{eq:phi_eigv_cons}
\end{align}
For the five Bravais lattices with n.n.~hopping and either four or six n.n.~neighbors the above generic expressions can be further simplified. The case of four nearest neighbors---square, oblique, and rectangular lattices---the real space wavefunction is given by products of pairs of Bessel functions:
\begin{align}
    \ket{\Phi} =  \sum_{(n,m)\in\mz^2}~ J_{m}\left(\frac{2 s_{0,1}}{\ya}\right)  J_{n}\left(\frac{2 s_{1,0}}{\xa}\right)\ket{n,m}.
    \label{eq:square_func}
\end{align}
The eigenfunctions on triangular and centered rectangular lattices are given by sums over products of triplets of Bessel functions:
\begin{align}
    \ket{\Phi} = &\sum_{(n,m)\in\mz^2}~ 
    \sum_{\nu\in\mz} J_\nu \left(-\frac{2 s_{1,-1}}{\xa - \ya} \right) \notag\\
    & \times J_{m - \nu}\left(\frac{2 s_{0,1}}{\ya} \right) J_{n + \nu}\left(\frac{2 s_{1,0}}{\xa} \right)\ket{n, m}.
    \label{eq:triangular_func}
\end{align}
The details of the derivations are given in Appendix~\ref{app:eig-46}.

We are interested in the decay properties of the above wave functions along and perpendicular to the field direction. Recall the asymptotics of Bessel functions $J_\nu(t) \sim \frac{1}{|\nu|!} \left|\frac{t}{2}\right|^{|\nu|}$ for large order integer $|\nu|$. Since the above wave functions involve products of Bessel functions, we conclude that the spatial decay will be at least factorial $1/r!$ in any lattice direction, which is faster than any exponential decay. Let us consider the square lattice with $s=t/\mathcal{F}$ and $t$ the nearest hopping strength. From Eq.~\eqref{eq:DZDW} it follows
\begin{align}
    \Phi(z = 0, w = \zeta \Delta_w) = J_{-\zeta x}\left(\frac{2 s}{y} \right) J_{\zeta y} \left(\frac{2 s}{x} \right), \\
    \Phi(z = \zeta \Delta_z, w = 0) = J_{\zeta y}\left(\frac{2 s}{y} \right) J_{\zeta x} \left(\frac{2 s}{x} \right),
\end{align}
with $\zeta \in \mz$. We therefore arrive at the wavefunction asymptotics for $|\zeta| \rightarrow \infty$
\begin{align}
    |\Phi(z = 0, w = \zeta \Delta_w)| \approx \frac{1}{|\zeta x|! |\zeta y|!} \left|\frac{|s|^{|x|+|y|}}{|y|^{|x|} |x|^{|y|}}\right|^{|\zeta|}, \\
    |\Phi(z = \zeta \Delta_z, w = 0)| \approx \frac{1}{|\zeta x|! |\zeta y|!} \left|\frac{|s|^{|x|+|y|}}{|y|^{|y|} |x|^{|x|}}\right|^{|\zeta|}, \\
    \frac{ \left|\Phi\left(z = 0, w = \zeta \Delta_w \right)\right|}{\left|\Phi\left(z = \zeta \Delta_z, w = 0\right)\right|} \approx \left(\left|\frac{x}{y}\right|^{|x| - |y|} \right)^{|\zeta|}. 
\end{align} 
Since $\Delta_w=\Delta_z$ and for nonzero integers $|x|\neq |y|$ the term $\left|x/y\right|^{|x| - |y|} > 1$, the flatband eigenstates always decay faster along the field direction as compared to the perpendicular one. The only exception is $|x|=|y|=1$, for which the decay in both directions is the same. 

\begin{figure}[h]
    \subfigure[]{\includegraphics[width=0.95\columnwidth]{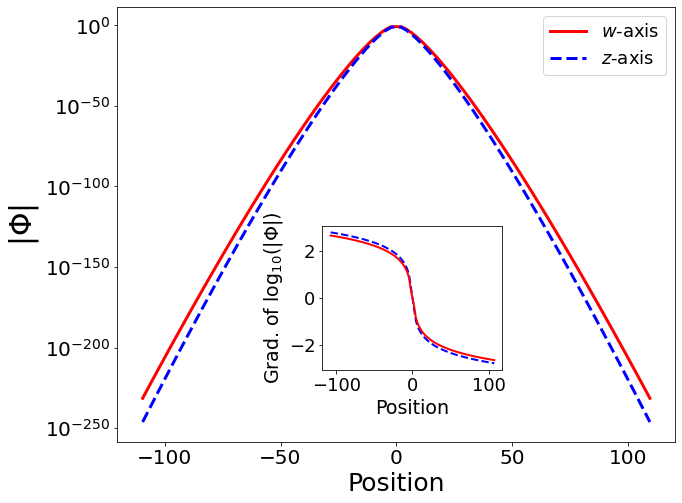}} 
    \subfigure[]{\includegraphics[width=0.95\columnwidth]{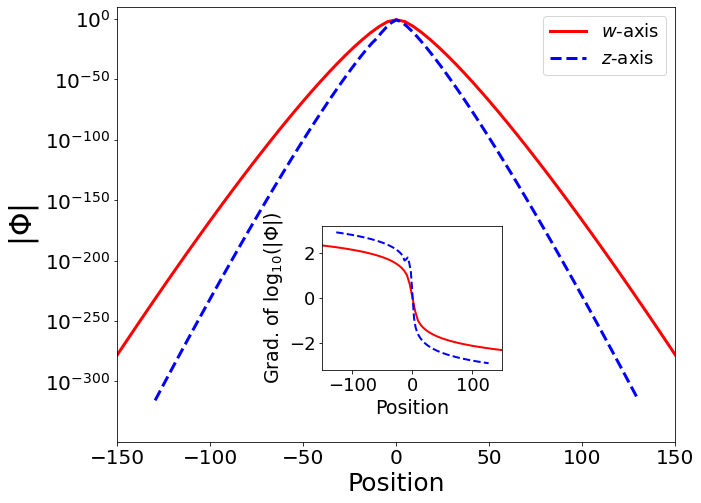}}
    \caption{Absolute values of the probability amplitude $|\Phi(z,w)|$ for (a) square lattice (b) triangular lattice---as a function of $z$ or $w$ positions in log-linear scale. The static field is along $(x,y) = (2,1)$ direction. Insets are the gradients of $\log_{10}(|\Phi(z,w)|)$ with respect to $z$ or $w$. The position ticks are scaled according to the physical distances.}
    \label{fig:func_decay} 
\end{figure}

We plot the wave function profile for the square lattice and field direction $x=2,y=1$ in Fig.~\ref{fig:func_decay}{\color{red}(a)} in log-linear scale. The wavefunction decays faster along the field direction than in the perpendicular one. The inset shows that the gradient is monotonically decreasing as a function of position for both directions, i.e., the decay is faster than exponential (for exponential decay it would be a step-function around the origin).

We analyzed numerically a similar case for the triangular lattice, which is shown in Fig.~\ref{fig:func_decay}{\color{red}(b)}. The field direction is again $x=2,y=1$. Again we observe that the wavefunction is decaying faster along the field direction than in the perpendicular one, and in both directions faster than exponential (see inset). We therefore expect that other Bravais lattices and commensurate field directions will yield similar localization properties of flatband eigenstates. It appears reasonable that the decay in the perpendicular direction is slower than in the field  direction, as the onsite energies of the original lattice vary linearly with distance along the field direction as opposed to the perpendicular one.

\section{Dimension $d \geq 3$}
\label{higher_dim}

It is straightforward to observe that we can define commensurate field directions for higher $d$-dimensional Bravais lattice in the same way: Choose any two points on the lattice separated by a finite distance, then the connecting line corresponds to an allowed perpendicular direction. In this case the quasimomentum along the perpendicular direction $k$ will be replaced by a $d-1$ dimensional vector $\vec{k}$ in the generating function in Eq.~\eqref{eq:GF}, keeping the $2\pi$ periodicity of the generating function with respect to $q$: $g_\sem(q + 2\pi, \vec{k})$ = $g_\sem(q, \vec{k})$. The differential equation [Eq.~\eqref{eq:diff_g}] will remain similar after replacing $k$ by $\vec{k}$. In the absence of hopping connections along the perpendicular direction, the solution of the differential equation [Eq.~\eqref{eq:diff_g}] turns
\begin{align}
    g_\sem(q, \vec{k}) = A(\vec{k}) e^{-i\lambda_\sem q} e^{-i f(q, \vec{k})},
\end{align}
where $A(\vec{k})$ and $f(q, \vec{k})$ are periodic functions of $\vec{k}$ and $q$, and $\lambda_\sem = E/\mathcal{F}$. Therefore, similar to the 2D case, the periodicity condition $g_\sem(q + 2\pi, \vec{k}) = g_\sem(q, \vec{k})$ implies $\lambda_\sem$ takes all possible integer values, and hence the entire spectrum degenerates into an infinite set of WS flatbands. In the presence of hopping connections along the chosen perpendicular direction, the bands will acquire nonzero dispersion.

For all commensurate field directions each band is characterized by $(d-1)$-dimensional wavevectors $\vec{k}$, while the wavefunctions are embedded into a $d$-dimensional space. We expect that flatband eigenstates will be localizing faster than exponential in all directions, and slower in the perpendicular directions as compared to the field direction.

As an example, we consider the case of the cubic lattice with nearest-neighbor hopping $t$ that respects the inversion symmetry, Eq.~\eqref{eq:inv_sym}. The Hamiltonian then is a 3D version of Eq.~\eqref{eq:tb-ham}. The derivation of the generating function, and the spectrum in this case, is a straightforward generalization of that for the square lattice. Therefore, we only provide the final results. For a commensurate field direction $\vec{\mE} \propto F(x_1, x_2, x_3)^T$ with $x_1 x_2 x_3 \neq 0$ and gcd($x_1, x_2$) = gcd($x_1, x_3$) = 1, the generating function reads ($\vec{k} = (k_1,k_2)$),
\begin{gather*}
    g_\sem(q, \vec{k}) = A(\vec{k}) \exp\Bigg[-i \lambda_\sem q -i \sum_{(j_1,j_2,j_3) \in \mathcal{R}_3} 2s \\
    \times \frac{\sin\left[q(j_1 x_2 + j_2 x_2 + j_3 x_3) + k_1 \epsilon_1 + k_2 \epsilon_2 \right]} {j_1 x_2 + j_2 x_2 + j_3 x_3}\Bigg], \\
    \mathcal{R}_3 = \{(1, 0, 0), (0, 1, 0), (0, 0, 1)\},\\
    \epsilon_1 = j_3 - (j_1 x_1 + j_2 x_2 + j_3 x_3),~ 
    \epsilon_2 = j_1 \tau_1 - j_2 \tau_2,
\end{gather*}
where $\lambda_\sem = E/\mf$, $s=t/\mf$, and
$\tau_2 x_1 + \tau_1 x_2 = 1$ for fixed integer values of $x_1$, $x_2$. Similarly, to the 2D case the $2\pi$ periodicity in $q$ of the generating function fixes the eigenenergies and implies that the spectrum consists of 2D flatbands $E(\vec{k}) = a\mf$ with $a\in\mz$. The inverse Fourier transform of $g_\sem$ yields the eigenfunction ($a = 0$),
\begin{gather*}
    \ket{\Phi} = \sum_{(n, m, l)\in\mz^3} J_{n}\left(\frac{2s}{x_1}\right) J_{m}\left(\frac{2s}{x_2}\right) J_{l}\left(\frac{2s}{x_3}\right)\ket{n,m,l},
\end{gather*}
that clearly displays factorial decay in the lattice coordinates $(n,m,l)$.

\section{Discussion and conclusions}

We considered the impact of a dc field on the spectra of tight-binding models on $d$-dimensional Bravais lattices. For commensurate field directions (for which the perpendicular direction is parallel to some lattice vector) the spectrum consists of an infinite number of equidistant $(d-1)$-dimensional bands. A finite number of commensurate directions yields dispersive bands. The remaining infinite set of commensurate directions leads to dispersionless Wannier-Stark flatbands. The flatband wavefunctions are embedded in a $d$-dimensional vector space, and cease to form compact localized states, which is the usual scenario for translationally invariant lattices with short-range hoppings. As a result, Wannier-Stark flatband eigenfunctions decay factorially in space, and typically faster along the field direction than perpendicular to it.

Our results are applicable for ultracold atoms in optical lattices, where the electric field is substituted by a tilt of the lattice in the gravitational field~\cite{anderson1998macroscopic} or acceleration of the whole lattice~\cite{christiani2002experimental}. Notably, the same type of perturbations can be arranged  in optical waveguide arrays where the electric field is modeled by a curved geometry of the waveguides~\cite{longhi2006observation}. In both cases experimental platforms for two-dimensional settings have been developed. Such experiments can test the sensitivity of choosing commensurate field directions, the existence of flatband field directions, Bloch oscillations, and the factorial localization of flatband eigenstates. Another intriguing set of potential applications could be related to electronic transport in strained two-dimensional materials, which is, e.g., a hot topic in graphene research~\cite{naumis2017electronic}.

Some open problems for Wannier-Stark flatbands on Bravais lattices are their fate in the presence of perturbations such as disorder, magnetic fields, and many-body interactions. The method used in this work might not be applicable in these cases, and one would need to resort to the generic analytical and numerical methods. However, a possible way to analyze the impact of the perturbations is to perform perturbation calculations around the limit of strong dc field strength, which allows the approximation of the factorially localized eigenstates by compact ones and use of the methods developed for the analysis of disordered and/or interacting flatband models~\cite{danieli2020nonlinear,danieli2020quantum,leykam2013flat,flach2014detangling}. An even more interesting problem is the case of non-Bravais lattices where there is more than one site per unit cell. The generating function approach extends naturally to non-Bravais lattices~\cite{kolovsky2018topological} and again leads to a Wannier-Stark ladder of an irreducible band structure, now consisting of several bands. Can some of these bands be tuned to become flat? A positive answer exists for some chiral lattices, which transport the chiral symmetry into the generating function and the irreducible band structure~\cite{kolovsky2018topological}. Similar to translationally invariant chiral flatbands~\cite{ramachandran2017chiral}, the irreducible Wannier-Stark band structure will contain a chiral flatband. The eigenfunctions will be noncompact as in the Bravais lattice case. At variance to the Bravais case they were observed to localize only exponentially~\cite{kolovsky2018topological}, perhaps due to the presence of other dispersive nonflat bands in the irreducible spectrum. Can we finetune non-Bravais lattice hoppings such that the irreducible Wannier-Stark band structure turns one, or several, or even all, bands flat---without imposing a symmetry like the chiral one? We think these are exciting questions for future research.

\begin{acknowledgments}
    This work was supported by IBS-R024-D1. N.C.~would like to thank the China Scholarship Council (CSC-201906040021) for financial support.  
\end{acknowledgments}
\appendix
\section{Commensurate field directions}
\label{app:derivation_comm}
A commensurate field direction is defined by requiring that the direction perpendicular to the dc field,
\begin{gather*}
    \vec{n}_\perp \propto -y \gamma_1 \sin \theta \hat{e}_0 + (x \gamma_0 + y \gamma_1 \cos \theta) \hat{e}_1,
\end{gather*}
is parallel to some lattice vector indexed by $p_1, p_2 \in \mz$:
\begin{gather*}
    p_1\ha_0 + p_2\ha_1 = (p_1 \gamma_0 + p_2 \gamma_1 \cos \theta) \hat{e}_0 + p_2 \gamma_1 \sin \theta \hat{e}_1.
\end{gather*}
The condition of these two vectors being parallel reads
\begin{gather*}
    \frac{(p_1 \gamma_0 + p_2 \gamma_1 \cos \theta)}{-y \gamma_1 \sin \theta} = \frac{p_2 \gamma_1 \sin \theta}{x \gamma_0 + y \gamma_1 \cos \theta},
\end{gather*}
and can be rewritten as
\begin{gather*}
    (p_1 x \gamma_0^2 + p_2 y \gamma_1^2) = -(p_1 y  + p_2 x)\gamma_0 \gamma_1 \cos \theta.
\end{gather*}
This implies that either of the following conditions on $x$ and $y$ hold:
\begin{itemize}
    \item[(1)] $p_1 y + p_2 x = p_2 y \gamma_1^2 + p_1 x \gamma_0^2 = 0$ implying
    \begin{gather*}
        \frac{p_1}{p_2} = -\frac{x}{y} = - \frac{y \gamma_1^2}{x \gamma_0^2}.
    \end{gather*}
    From the above it follows that both $\frac{x}{y}$ and $\frac{\gamma_1}{\gamma_0}$ are rational, and $|x \gamma_0| = |y \gamma_1|$. 
    \item[(2)
    ] $p_1 y + p_2 x \neq 0$ implying
    \begin{gather*}
        \cos \theta = -\frac{p_2 y \gamma_1^2 + p_1 x \gamma_0^2}{(p_1 y + p_2 x) \gamma_0 \gamma_1}.
    \end{gather*}
\end{itemize}

\section{Parametrization of the rotated coordinates $z,w$ in 2D}
\label{app:para}

In the main text, Eq.~\eqref{eq:zw-def}, we defined our new coordinates as
\begin{align}
    z = n \xa + m \ya, ~ w = ny - mx,
\end{align}
where $\xa$ and $\ya$ were defined as 
\begin{align}
    \xa = \alpha(x\gamma_0^2 + y \gamma_0 \gamma_1 \cos \theta), \notag\\
    \ya = \alpha(x \gamma_0 \gamma_1 \cos \theta + y \gamma_1^2).
    \label{eq:xalpha_yalpha}
\end{align}
The conditions for a commensurate field direction given by Eq.~\eqref{eq:comm_condition} imply the existence of a rescaling parameter $\alpha \neq 0$ for which $\xa$ and $\ya$ are integers, i.e., either $\xa \ya = 0$ or $\xa/\ya$ is rational. 

Plugging the first condition given by Eq.~\eqref{eq:comm_condition} in Eq.~\eqref{eq:xalpha_yalpha} we get
\begin{align}
    \xa = \alpha x \gamma_0^2 (1 \pm \cos \theta),~ & \ya = \alpha y \gamma_1 (1 \pm \cos \theta)\notag\\
    \Rightarrow \frac{\xa}{\ya} = & \pm \frac{\gamma_0}{\gamma_1},
\end{align}
therefore $\frac{\xa}{\ya}$ is rational since $\frac{\gamma_0}{\gamma_1}$ is rational. 

Putting the second condition of Eq.~\eqref{eq:comm_condition}) in Eq.~\eqref{eq:xalpha_yalpha} we find
\begin{align*}
    & x_\alpha = \alpha\left[x \gamma_0^2  - y\frac{p_2 y \gamma_1^2 + p_1 x \gamma_0^2}{(p_1 y + p_2 x)}\right], \\
    & y_\alpha = \alpha\left[y \gamma_1^2 - x\frac{p_2 y \gamma_1^2 + p_1 x \gamma_0^2}{(p_1 y + p_2 x)} \right]. 
\end{align*}
After some simple algebra one arrives at the required result:
\begin{gather*}
    \frac{x_\alpha}{y_\alpha} = - \frac{p_2}{p_1}
\end{gather*}

For $\xa \ya \neq 0$ we absorb gcd$(\xa, \ya)$ in $\alpha$ to make $\xa$, $\ya$ coprime for the convenience of our analysis. In the case $x_\alpha y_\alpha = 0$, we can choose either $\xa = 1$ while $\ya = 0$ or $\ya = 1$ while $\xa = 0$.

Therefore $z$ is integer for all $(n,m) \in \mz \times \mz$. On the other hand, $w$ takes discrete values, that are not necessarily integer or can not be made integer by rescaling of $w = ny - mx$ by a constant factor, since $x/y$ is not a rational number in general. Let us denote the set of all possible pairs $(z,w)$ by $\ms$ and construct its parametrization. We show that $z$ can take any integer value, and we can parameterize $w$ for a fixed value of $z$. As mentioned before, either $x_\alpha y_\alpha = 0$ (for which we can make either $x_\alpha = 1$ or $y_\alpha = 1$) or $x_\alpha$ and $y_\alpha$ are mutually prime. For fixed $z$, we pick one lattice point that corresponds to $z$, $w_0(z)$, and is indexed by $(n_0, m_0)$   
\begin{align}
    n_0 = \tau_2 \lambda,\quad m_0 = \tau_1 \lambda, 
    \label{fix_pt}
\end{align}
with $\tau_1$, $\tau_2$, $\lambda$ $\in$ $\mz$, 
\begin{align}
    z = \lambda(\tau_1 y_\alpha + \tau_2 x_\alpha).
\end{align}
According to B\'ezout's identity from number theory~\cite{jones2012elementary}, there is always a choice of $\tau_1$, $\tau_2$ such that 
\begin{align}
    \tau_1 y_\alpha + \tau_2 x_\alpha = 1.
    \label{eq:ident_rel}
\end{align}
This implies $z = \lambda$, and it takes any integer value. The values of $\tau_1, \tau_2$ can be determined for example using the Euclidean division algorithm \cite{ferguson1999analysis}. Note that despite the fact that the choice of $\tau_1$, $\tau_2$ is not unique, the condition (\ref{eq:ident_rel}) remains unchanged under the simultaneous change of $\tau_1$ by $\tau_1 - p \xa$ and $\tau_2$ by $\tau_2 + p y_\alpha$, where $p$ can be any integer. The corresponding value of the perpendicular coordinate is
\begin{align}
    w_0(z) = n_0 y - m_0 x =  \lambda (\tau_2 y -  \tau_1 x).
\end{align}
For a given integer value of $z$ all the other values of $w$ are generated by simultaneous shifts of $n$ from $n_0$ and $m$ from $m_0$ by the following way:
\begin{align}
    & n = n_0 + \eta \ya,\quad m = m_0 - \eta \xa,\quad\eta\in\mz, \notag\\
    & w = ny - mx = w_0(z) + \eta(\ya y + \xa x).
\end{align}
Therefore the entire set $\ms$ of valid lattice points is parameterized as~\cite{kolovksy2012cyclotron}
\begin{align}
    (z, w) = (z, w_0(z) + \eta(x \xa + y \ya)),~\quad z, \eta \in \mz.
    \label{eq:z_w_val}
\end{align} 
We use the parametrization~\eqref{eq:z_w_val} in the main text. In the following we provide the values of parameters  $\tau_1, \tau_2$ for some special cases. 
\begin{itemize}
    \item[(i)] if $\ya = 0, \xa = 1$, then $\tau_1 = 0, \tau_2 = 1$.
    \item[(ii)] if $\xa = 0, \ya = 1$, then $\tau_2 = 0, \tau_1 = 1$.
    \item[(iii)] if $\xa = \pm\ya, \ya = 1$, then $\tau_2 = \pm 1, \tau_1 = 0$.
\end{itemize}

For some simple cases we can even guess restrictions on the set $\ms$. For example, when $x$, $y$ are integers the set $\ms$ is only a subset of $\mz \times \mz$ and does not contain all of its elements. One can check that by simply looking at the square lattice case, where $z = nx + my$, $w = ny - mx$. In that case for the field direction $x = 2$, $y = 1$ the point $(z, w) = (0,1)$ does not exist in the original lattice since it corresponds to fractional indices $(n,m) = (1/5, -2/5)$.

\section{The action of $\mh$ on basis states $\ket{\phi(z,k)}$ for the 2D system}
\label{app:action_of_ham}

We use the identities:
\begin{align*}
    \Delta w \coloneqq (x \xa + y \ya),~ \tau_1 \ya + \tau_2 \xa = 1.
\end{align*}
The action of the Hamiltonian on the basis vector is given by: 
\begin{align*}
    \mathcal{H}\ket{\phi(z,k)} = \mathcal{F} z  \ket{\phi(z,k)} - \sum_{lj} t_{lj} \ket{z - lx_\alpha - j y_\alpha} \notag \\
    \otimes \sum_\eta e^{ik\eta} \ket{w_0(z) + \eta \Delta w - ly + jx}.
\end{align*}
In the right-hand side of the above equation the $w$ coordinate value is
\begin{align*}
    & w_0(z) + \eta \Delta w - ly + jx \notag \\
    =~  & \lambda(\tau_2 y - \tau_1 x) + \eta \Delta w - ly + jx \notag \\
    =~  & (\lambda - l\xa - j \ya)(\tau_2 y - \tau_1 x) \notag \\
    & +(l\xa + j \ya)(\tau_2 y - \tau_1 x) + \eta \Delta w - ly + jx \notag \\
    =~  & w_0(z - lx_\alpha - j y_\alpha) + \eta'\Delta w,
\end{align*}
with 
\begin{align*}
    \eta' =~ & \eta + \frac{l}{\Delta w}[ \tau_2 y \xa - \tau_1 x \xa - y] \notag \\
    & + \frac{j}{\Delta w}[\tau_2 y \ya - \tau_1 x \ya + x] \notag \\
    =~ & \eta - \tau_1 l + \tau_2 j = \eta - \epsilon_{lj},
\end{align*}
where we have defined
\begin{align*}
    \epsilon_{lj} = \tau_1 l - \tau_2 j.
\end{align*}
Then 
\begin{align*}
    & \sum_{lj} t_{lj} \ket{z - lx_\alpha - j y_\alpha} \otimes \sum_\eta e^{ik\eta} \ket{w_0(z) + \eta \Delta w - ly + jx}  \notag \\
    = & \sum_{lj} t_{lj} \ket{z - lx_\alpha - j y_\alpha} \notag \\
    & \otimes \sum_{\eta'} e^{ik\eta'} e^{-ik(\tau_2 j - \tau_1 l)} 
    \ket{w_0(z- lx_\alpha - j y_\alpha) + \eta' \Delta w} \notag \\
    = & \sum_{lj} t_{lj} e^{ik \epsilon_{lj}} 
    \ket{\phi(z- lx_\alpha - j y_\alpha, k)}.
\end{align*}
Therefore,
\begin{align*}
    \mathcal{H}\ket{\phi(z,k)} = \mathcal{F} z \ket{\phi(z,k)} \notag \\
    - \sum_{l,j} t_{lj} e^{ik \epsilon_{lj}} \ket{\phi(z- lx_\alpha - j y_\alpha, k)}.
\end{align*}
The basis states $\bra{\phi(z',k')}\ket{\phi(z, k)} = 2\pi \delta_{z,z'} \delta(k - k')$ form an orthonormal set and a complete basis for the Hilbert space on which the Hamiltonian operates.  

\section{Derivation of the Bessel function order dependence on the spatial coordinates in Eqs.~(\ref{eq:square_func}) and (\ref{eq:triangular_func})}
\label{app:eig-46}

Let us consider the case of triangular and centered rectangular lattices with six nearest neighbor hoppings. The case of square, rectangular, and oblique lattices with four nearest neighbor hoppings follows straightforwardly by setting $\nu = \nu_{(1,-1)} = 0$, $s_{1,-1} = 0$ in Eq.~\eqref{eq:triangular_func}. The constraints in Eq.~\eqref{eq:phi_eigv_cons} can be expanded as follows:
\begin{align*}
    & \nu_{(1,0)} \xa + \nu_{(0,1)} \ya + \nu_{(1,-1)}(\xa - \ya) = -(z-a), \notag \\
    & \nu_{(1,0)} \epsilon_{1,0} + \nu_{(0,1)} \epsilon_{0,1} + \nu_{(1,-1)} \epsilon_{1,-1} = -\eta,
\end{align*}
where $a$ is the band index which corresponds to different eigenenergies. Using the relations
\begin{align*}
    \tau_1 \ya + \tau_2 \xa = 1, \\
    \epsilon_{1,0} = \tau_1, \epsilon_{0,1} = - \tau_2,~\epsilon_{1,-1} = \epsilon_{1,0} - \epsilon_{0,1}, \\
    \eta = \frac{w - w_0(z)}{x \xa + y \ya} 
    = \frac{w - z[\tau_2 y  - \tau_1 x]}{x \xa + y \ya}, \\
    n = \frac{xz + w\ya}{x\xa + y\ya},~m = \frac{yz - w\xa}{x\xa + y\ya},
\end{align*}
we get
\begin{align*}
    \nu_{(0,1)} &= -z\tau_1 + \eta \xa +  \nu_{(1,-1)} + a \tau_1 \notag \\
    &= \nu_{(1,-1)} - m + a \tau_1; \notag \\
    \nu_{(1,0)} &= -z\tau_2  - \eta \ya -  \nu_{(1,-1)} 
    + a \tau_2 \notag \\
    &= - n -\nu_{(1,-1)}+ a \tau_2.
\end{align*}
In the main text we analyzed the $a = 0$ case only. We also used the following symmetry properties of the Bessel functions:
\begin{align*}
    J_{-\nu}(-\mu) =  J_{\nu}(\mu),
\end{align*}
for integer order index $\nu$ and real $\mu$. 

\section{Nonexistence of compact localized eigenstates for Wannier-Stark flatbands}
\label{app:cls_nonproof}

Let us try to construct a compact localized eigenstate $\ket{\Phi_\text{CLS}}$ in position space, assuming existence of a set of proper superposition coefficients $C_\text{CLS}(k)$:
\begin{align*}
    \ket{\Phi_\text{CLS}} = \frac{1}{2\pi}\int_{k} C_\text{CLS}(k) \ket{\psi(k)} dk \notag \\
    = \frac{1}{4\pi^2} \int_{k}\int_{q} dq dk ~C_\text{CLS}(k) \notag \\
    \times \sum_{z, \eta \in \mz} e^{i(qz + k\eta)}g_\sem(q,k) \ket{z, w_0(z) + \eta \Delta w}. 
\end{align*}
The generating function $g_\sem(q,k)$ is periodic both in $k$ and $q$, and hence it can be  expanded as a Fourier series in the variables $k$ and $q$:   
\begin{align*}
    g_\sem(q,k) = \sum_{p_1, p_2 \in \mz} g_{p_1, p_2} e^{i k p_1+ i q p_2}. \notag
\end{align*}
Therefore 
\begin{align*}
    \ket{\Phi_\text{CLS}} & = \frac{1}{2\pi}\int_{k} dk ~C_\text{CLS}(k)\times \notag \\
    \sum_{p_1, p_2, \eta \in \mz} & e^{ik(\eta+p_1)} g_{p_1, p_2}\ket{-p_2, w_0(-p_2) + \eta \Delta w}.
\end{align*}
Compactness of the eigenstate in the field direction implies $g_{p_1,p_2} = 0$ except for a finite number of values of $p_2$. But from the solution Eq.~\eqref{eq:flat_g} it follows that the generating function cannot be expressed as finite polynomial in $e^{iq}$. Hence a flatband wavefunction cannot be made compact in the direction of the field.

We will now discuss the possibility for the flatband wavefunction to be compact in the perpendicular direction of the 
field. Since we study lattice eigenvalue problems, $C_\text{CLS}(k)$ is a $2\pi$-periodic function in $k$: 
\begin{align*}
    C_\text{CLS}(k + 2\pi) = C_\text{CLS}(k)\Rightarrow C_\text{CLS}(k)= \sum_{p_3 \in \mz} C_{p_3} e^{ikp_3}.
\end{align*}
Therefore,
\begin{align*}
  \ket{\Phi_\text{CLS}} =   &
    \sum_{p_1, p_2, p_3 \in \mz} C_{p_3}  g_{p_1, p_2}\notag\\
   & ~~~~~
    \ket{-p_2, w_0(-p_2) -(p_1 + p_3) \Delta w}.
\end{align*}
Compactness in the direction perpendicular to the field implies that the product $C_{p_3} g_{p_1, p_2} = 0$ except for a finite number of integer values of the sum $(p_1 + p_3)$. Simple inspection of Eq.~\eqref{eq:flat_g} yields that, for any fixed value of $p_3$ with a corresponding nonzero $C_{p_3}$, there always exists an infinite number of $p_1$ values  for which $g_{p_1, p_2}$ turns nonzero. 

Therefore there exists no function $C_\text{CLS}(k)$ for which the flatband eigenfunction turns into a CLS. Moreover, we proved that the flatband eigenfunctions are necessarily non-compact in all space directions. The proof can be generalized to higher space dimensions $d\geq 2$ by replacing $k$ with a $(d-1)$ dimensional vector $\vec{k}$.

\bibliography{general,flatband}

\end{document}